# First-principles study of hydrogen storage over Ni and Rh doped BN sheets

Natarajan Sathiyamoorthy Venkataramanan,[a *] Mohammad Khazaei,[a]
Ryoji Sahara,[a] Hiroshi Mizuseki,[a] Yoshiyuki Kawazoe[a]

[b]*Institute for Materials Research(IMR),*
*Tohoku University,*
*2-1-1 Katahira,*
*Aoba-ku,*
*Sendai 980-8577, Japan*

* Corresponding author: ramanan@imr.edu;   Phone : +81-22-215-2054;  Fax : +81-22-215-2052

**Abstract :**

Absorption of hydrogen molecules on Nickel and Rhodium – doped hexagonal boron nitride(BN) sheet is investigated by using the first principle method. The most stable site for the Ni atom was the on top side of nitrogen atom, while Rh atoms deservers a hollow site over the hexagonal BN sheet. The first hydrogen molecule was absorbed dissociatively over Rh atom, and molecularly on Ni doped BN sheet. Both Ni and Rh atoms are capable to absorb up to three hydrogen molecules chemically and the metal atom to BN sheet distance increases with the increase in the number of hydrogen molecules. Finally, our calculations offer explanation for the nature of bonding between the metal atom and the hydrogen molecules, which is due to the hybridization of metal d orbital with the hydrogen s orbital. These calculation results can be useful to understand the nature of interaction between the doped metal and the BN sheet, and their interaction with the hydrogen molecules.

## 1. Introduction :

Hydrogen is attractive as a fuel because its use creates neither air pollution nor greenhouse-gas emissions [1–3]. The use of hydrogen requires an effective, safe, and stable storage medium. However, how to store hydrogen easily and cheaply is still a big and challenging problem [4, 5]. The current methods of storing hydrogen as compressed gas or in the liquid form do not meet the industrial requirements because the energy densities are much lower than that in gasoline. Moreover, there are issues of safety and cost involved in compressing hydrogen under high pressure or liquefying it at cryogenic temperatures. Although storage of hydrogen in solid-state materials offers an alternative, there are no current solid-state storage materials that meet the industry requirements.

In the past, considerable attention has been focused on porous materials such as Metal-organic Frameworks (MOF's), clathrates, carbon nanotubes, and fullerenes as possible materials for hydrogen storage [6–10]. Recent efforts have been directed at non-carbon nanosystems composed of light elements such as B and N [11]. B-N nanostructures are an analogue of the carbon ones and offer many advantages. Carbon based materials are oxidized at $600^{o}$ C while B-N materials are stable upto $1000^{o}$ C. The heteropolar nature in the BN sheets offers higher binding energy for hydrogen compared to the carbon based materials. It has been found experimentally that at 10 MPa the BN nanotubes can store as much as 2.6 wt % of hydrogen and that collapsed BN nanotubes exhibit an even higher storage capacity (4.2 wt%) [12,13].

Boron nitride can exist in a hexagonal structure *h*-BN similar to the graphite sheet structure. Very recently Osterwalder et al. prepared BN sheets by high-temperature exposure of the clean rhodium surface to borazine [14]. Following the work, BN sheets were prepared by exposing borazine on the transition metal surface [15–17]. Hydrogen storage properties of BN nanotubes have been well investigated, however only little attention has been paid to sheets [18–22]. Alonso et al. using DFT method claimed that adsorption capacity of *h*-BN sheets is similar to carbon nanotubes [23]. Very recently, Shevlin and Guo found that binding properties of boron-nitiride systems are strongly dependent on the defects and dopants present on the sheets [24].

Doping of transition elements was found to increase the hydrogen storage properties of materials [25]. Especially nickel and rhodium are widely used in the hydrogenation reaction and also in the synthesis of BN sheets. It is worth to mention that BN sheet preparation involves transition metals unlike in carbon system, which needs external doping that cannot easily controlled in experiments. Considering the potential application of Ni and Rh nanoparticles in hydrogen storage and in catalysis, in this paper we computationally report the first attempt on the interaction of Ni and Rh atom on the BN sheets through the first-principles calculations. We also analyze the interaction between hydrogen molecules and the metal atoms adsorbed on the BN sheets, which might be useful to maximize the hydrogen storage capacity.

## 2. Computational details

All of our calculations have been carried out using Density Functional Theory (DFT) within the Generalized-Gradient Approximation (GGA), with the exchange–correlation functional of Perdew and Wang (PW91) [26]. We used ultrasoft pseudopotential as implemented in the Vienna *Ab initio* Simulation Package (VASP), which spans reciprocal space with a plane-wave basis with a kinetic energy cutoff of 400 eV . Spin polarized calculations were carried out and all Brillouin zone integrations were done at the gamma point. Periodic boundary conditions were imposed and each BN sheets consists of 30 atoms of B and 30 atoms of N and are separated by ca. 13.5 Å, and with a nearest neighbor distance of 1.51 Å. The structural and ionic relaxations are alternatively performed on the cell, until the atomic forces are converged to ±0.01 eV/Å. The absorption energy for hydrogen over the metal doped BN sheets was calculated using

$$E_{abs} = E(BN_S\text{+}M\text{+ } nH_2) - E(BN_S\text{+ } M) - E(nH_2)$$

Where $E(BN_s\text{+}M\text{+ } nH_2)$ is the total energy of BN sheets doped with metal and absorbed with hydrogen, $E(BN_S\text{+ } M)$ is the total energy of the BN sheet doped with metal atom, and $E(nH_2)$ is the total energy of hydrogen molecules.

## 3. Results

### *3.1. Position of metal atom on BN sheet and bonding*

We first explored the absorption site on which metal atom can give rise to the highest absorption energy on the BN sheet. Four possible sites, including the top site of the boron atom (B), the top site of the nitrogen atom (N), the bridge site over BN bond (BN) and the hollow site of the hexagon BN (H) ring as shown in **Fig.1(a)** were tested. After full structural optimization, the Ni atom was found to locate at either N or H site, regardless of the initial location of the Ni atom. The most stable structure for the Ni atom doped BN sheet was found to be the N site, while the H site is less stable by 2.76 eV. Indeed, Shin and co-worker, found that Ni atoms prefer the bridge site in the fullerenes [27]. When Ni atom is capped on the BN sheet, the B – N distance around the Ni atoms is slightly elongated from 1.53 to 1.57 Å, and the N – Ni distance was found to be 1.83 Å which is close to the value of 1.88 Å predicted by x-ray photoelectron diffraction(XPD) study [28]. For the Rh atom the best site was found to be on the hexagon BN ring, while the BN site was less stable by 2.92 eV. The distance between the nitrogen and Rh atom was 2.40 Å, whereas the distance between the boron and Rh atom was 2.33 Å.

To understand the bonding characteristics, we calculate the charge density difference of the system, which is defined as the difference between the total charge density and the atomic charge densities. The charge difference for the BN sheets with metal Ni and Rh are shown in **Fig. 2a** and **2b**. It is evident from the **Fig. 2a** and **Fig. 2b**, a noticeable chemical bonds were formed between Ni and N atom and Rh and BN hexagon respectively. In the case of Ni doped BN sheets, hybridization between Ni 3d orbital and 2p orbital of nitrogen was found to occur, while in the case of Rh doped BN sheets, 4d orbital of Rh mix with 2p orbital of boron and nitrogen. Thus a strong bonding between the doped metal atom and the BN sheets is evident from the nature of hybridization and the bond lengths between the metal atom and the BN sheet.

### *3.2. Hydrogen absorption over metal doped BN sheets.*

To examine the hydrogen-molecule adsorption on the metal doped BN sheets, we place one hydrogen molecule near the BN sheet. The optimized geometric structure of $H_2$ adsorbed on the Ni and Rh doped BN sheets are shown in **Fig. 3a** and **3b**. The binding energy per atom for the hydrogen molecule and the selected bond distance between H – H and N – M ( M= Ni, Rh) are provided in **Table 1**. The first hydrogen molecule is chemically adsorbed, which is evident from the high BE values per hydrogen molecule. In the case of Ni a partial dissociation of the H – H molecules was observed, whose length were increased from 0.75 Å to 0.88 Å and the computed $E_{abs}$ was found to be 1.41 eV. Further, the Ni – N distance got elongated to 1.88 Å, indicating hydrogen molecule absorption weakens the Ni – N bond. In the case of Rh atom, a complete dissociation was observed with H – H distance of 1.87 Å and BE value of 1.60 eV, with no appreciable change in the bond lengths between Rh- N and Rh –B. In the case of Pt doped BN tube, Wu and co-workers observed Pt – N bond strengthens, as it absorbs the first hydrogen molecule, with hydrogen chemisorbed over Pt atom [29].

We then investigate the addition of the second hydrogen molecule over the Ni and Rh doped BN sheets. Upon absorption of the second hydrogen molecule, $E_{abs}$ values decreases to 0.991 eV and 0.936 eV for Ni and Rh doped BN sheets respectively, while the H – H distance in Rh system drastically reduced to 0.92 Å with a slight elongation of Rh – N distance. In the case of Ni system no appreciable change in the H – H distance are observed; however a decrease in the $E_{abs}$ value along with increase in the Ni – N distance was noticed. Thus the second hydrogen molecule is chemically absorbed over the Ni and Rh system, along with a increase in the Rh – N bond length.

To know the maximum number of hydrogen molecules that can be chemically absorbed, we added third and fourth hydrogen molecules to the metal doped BN sheets. When the third hydrogen molecule is placed near the Ni atom, it was chemisorbed and was found to be at a distance of 1.61 Å from Ni atom, with a absorption energy of 0.739 eV. Furthermore, the Ni- N distance weakens further and the H – H distance for the third molecule was 0.85 Å. The lowest energy optimized geometry for the third hydrogen molecule adsorbed over Ni atom doped BN sheets is shown in **Fig. 3c** With the introduction of fourth hydrogen molecule near the Ni atom, the system gets optimized with one hydrogen molecule at 3.855 Å away from the Ni atom, whose H – H distance was 0.750 Å . Thus Ni atom can adsorb 3 molecules of hydrogen by chemisorption and the other hydrogen molecules stay in molecular form.

In the case of Rh doped BN sheets, the introduction of third hydrogen molecule was found to be chemisorbed over the Rh atom and was at a distance of 1.74 Å away from it and with a binding energy of 0.87 eV. However, the Rh atom was found to migrate from the hollow site of the hexagon BN ring to the on top site of the nitrogen atom and was 2.33 Å away from the nitrogen atom. The optimized geometry along with the three hydrogen molecules adsorbed over Rh atom doped BN sheet is provided in **Fig. 3d.** Upon introduction of fourth hydrogen molecule, the Rh atom was found to detach from the BN ring and was 3.60 Å away from the BN sheet. However the fourth hydrogen molecule was chemisorbed over the Rh atom and the detached Rh atoms acts as if in the cluster environment.

### *3.3. Nature of interaction between hydrogen and metal atom*

To understand the nature of binding between the metal atom and the hydrogen molecules, we calculate the partial density of states (PDOS). The calculated PDOS for Ni atom doped BN sheet and the hydrogen adsorbed Ni atom doped BN sheets are provided in **Fig. 4a** and **4b** respectively. From the **Fig. 4a** and **4b** it is visible that binding states just below the Fermi level $E-E_f$ is mainly attributed to the Ni d orbital hybridization, and simultaneously with the hydrogen s orbital after the hydrogen adsorption. The calculated PDOS for the Rh doped BN sheet and hydrogen adsorbed Rh doped BN sheet are shown in **Fig. 4c** and **4d**. In the case of Rh atom doped BN sheet, the d orbital of Rh are hybridized with p of boron and nitrogen atom. In the case of hydrogen adsorbed Rh doped BN sheet, Rh d orbital hybridized with the p orbital of boron and nitrogen and simultaneously with the s orbital of hydrogen. When comparing the TDOS structure for the Rh doped BN sheet and hydrogen adsorbed Rh doped BN sheet a considerable change in the structure is visible which we attribute to the hydride formation.

**4. Discussion :**

In the past several studies were carried out to know the adsorption of hydrogen molecule over various metal doped carbon, BN nanotubes and fullerenes [27, 29, 31, 32]. Zhao and co-workers suggested that BN bridge site to be the stable site for the Ni atom in BN nanotubes [33] . Similar results were observed by Wu and co-workers in the case of Pt doped BN nanotubes [29]. Our results suggest, for Ni atom doped over BN sheet, on top site is more stable by 2.76 eV than the hollow hexagonal site, while for the Rh atom doped BN sheet, the hollow site on the hexagon is more stable site. Charge density difference between metal atom and the BN sheet shows that Ni atoms are bound to nitrogen atoms through hybridization of Ni d orbital and the nitrogen p orbital, while in the case of Rh doped BN sheets, 4d orbital of Rh mix with 2p orbital of boron and nitrogen. Thus a strong bonding between the doped metal atom and the BN sheets is evident from the nature of hybridization and the bond lengths between the metal atom and the BN sheet.

Hydrogen adsorption studies over the metal doped BN sheets shows that both Ni and Rh atoms can hold three hydrogen molecules. In the case of Ni doped BN sheet all three hydrogen molecules are chemically bound and are not dissociate, while in the case of Rh doped BN sheets, the first hydrogen molecule was chemically dissociate and rest hydrogen molecule are chemically bound to the metal atom. The absorption energy for the first hydrogen was found to have large value for the Rh atom doped BN sheet, whereas for the second and third hydrogen molecules, the $E_{abs}$ are higher for the Ni doped BN sheets. Upon addition of fourth hydrogen molecule to the Ni doped BN sheets, the fourth hydrogen molecule is moved away to a distance of 3.855 Å. While in the case of Rh atom doped BN sheet, the Rh atom gets detached and acts similar to a cluster system. Thus Ni atoms are more stable on BN sheets, and have higher absorption energy compared to the Rh doped BN sheets. PDOS results shows that Ni d orbital hybridize with the hydrogen s orbital during bonding, and in the case of Rh doped BN sheets, Rh d orbital hybridized with the p orbital of boron and nitrogen and simultaneously with the s orbital of hydrogen.

## 5. Conclusion

In summary, we have performed first-principles calculations to understand the favorable absorption site of Ni and Rh atom doping on the BN sheets and their hydrogen adsorption properties. It was found that Ni atom prefers a on top site over nitrogen, while Rh atoms deservers a hollow site over the hexagonal BN sheet. A strong chemical bond is formed between the nitrogen and Ni atom in Ni doped BN sheet, while in Rh doped BN sheet, a mixing of boron and nitrogen bonds with Rh atom was found to occur. The first hydrogen molecule was absorbed dissociatively over Rh atom, and molecularly on Ni doped BN sheet. Both Ni and Rh atoms are capable to absorb up to three hydrogen molecules chemically and the metal atom to BN sheet distance increases with the increase in the number of hydrogen molecules. Finally, our calculations offer explanation for the nature of bonding between the metal atom and the hydrogen molecules, which is due to the hybridization of metal d orbital with the hydrogen s orbital.

## Acknowledgements

This work has been supported by New Energy and Industrial Technology Development Organization (NEDO) under "Advanced Fundamental Research Project on Hydrogen Storage Materials". The authors thank the crew of the Center for Computational Materials Science at Institute for Materials Research, Tohoku University, for their continuous support of the HITACHI SR11000 supercomputing facility.

**Legends:**

**Figure 1.** (a) The optimized geometric structure of BN sheets along with the possible site of metal doping (b) The optimized and most stable structure of Ni atom doped on BN sheet (c) The optimized and most stable structure of Rh atom doped on BN sheet.

**Figure 2.** Charge density difference for the metal atom doped on BN sheets (a) Ni atom doped on BN sheet (b) Rh atom doped on BN sheet.

**Figure 3.** The optimized geometric structures for the hydrogen adsorbed on the metal doped BN sheet (a) one hydrogen molecule adsorbed over Ni doped BN sheet (b) one hydrogen molecule adbsorbed over Rh doped BN sheet (c) three hydrogen molecules adsorbed over Ni doped BN sheet (d) three hydrogen molecules adsorbed over Rh doped BN sheet.

**Figure 4.** The Projected DOS of (a) Ni doped BN sheet (b) Ni doped BN sheet with one adsorbed hydrogen molecule (c) Rh doped BN sheet (d) Rh doped BN sheet with one adsorbed hydrogen molecule. The Fermi level is set to zero.

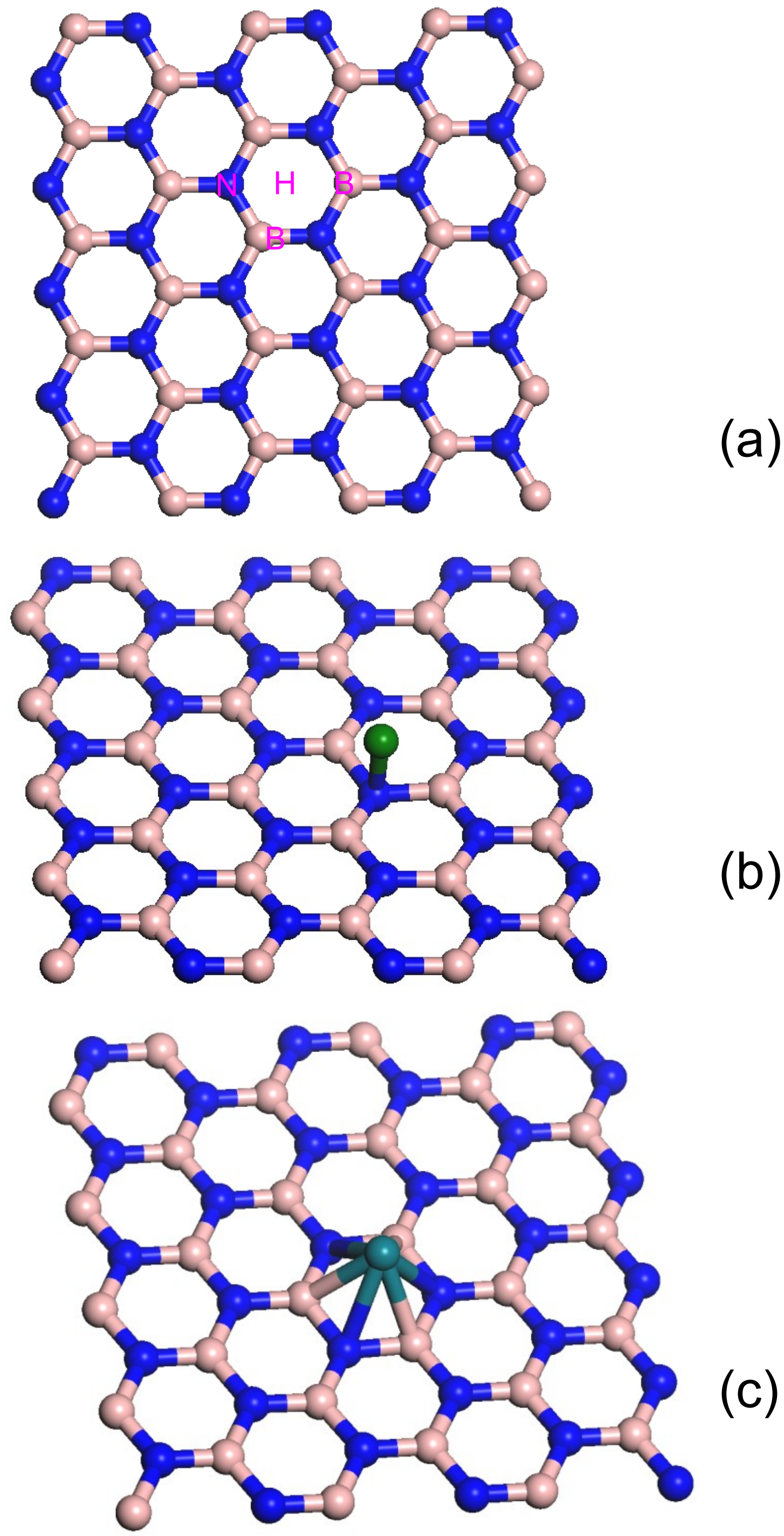

**Figure 1.**

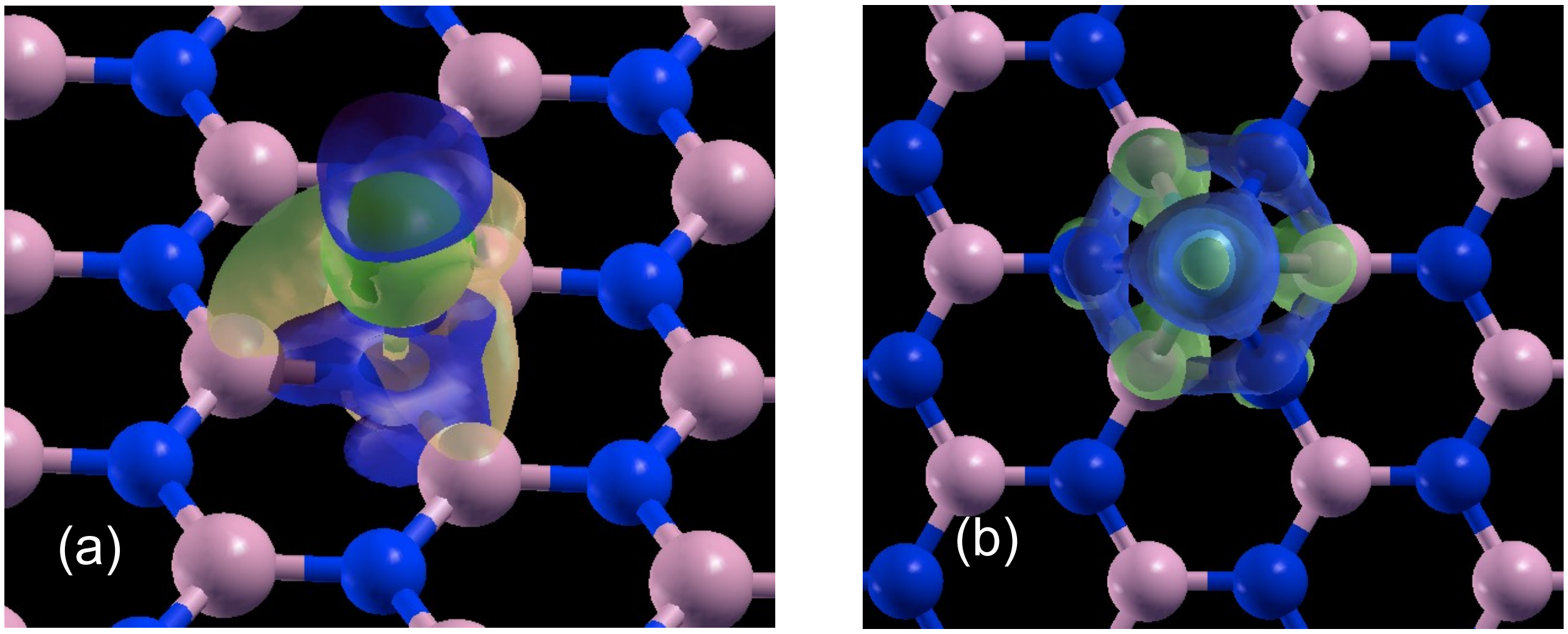

Figure 2

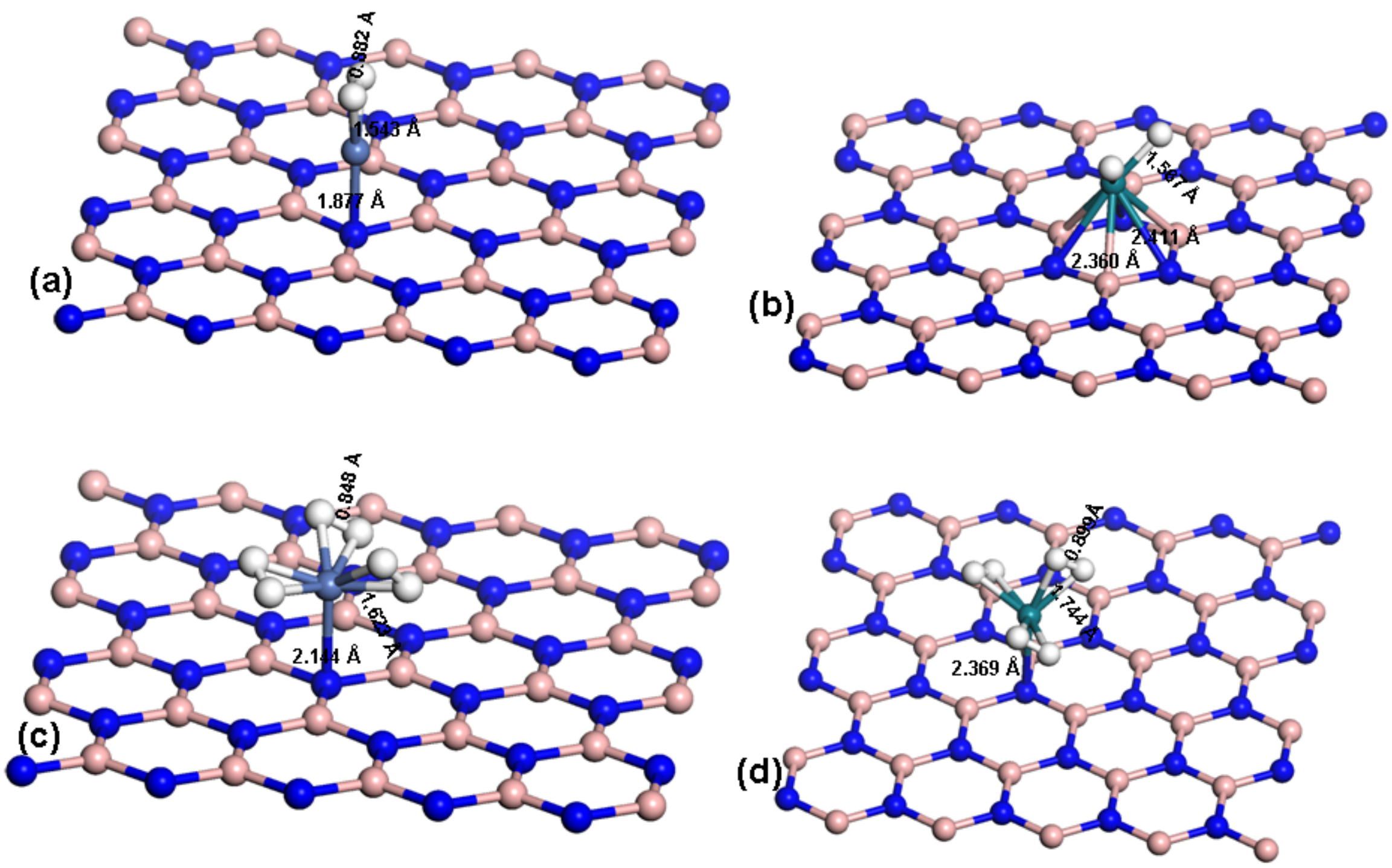

Figure 3

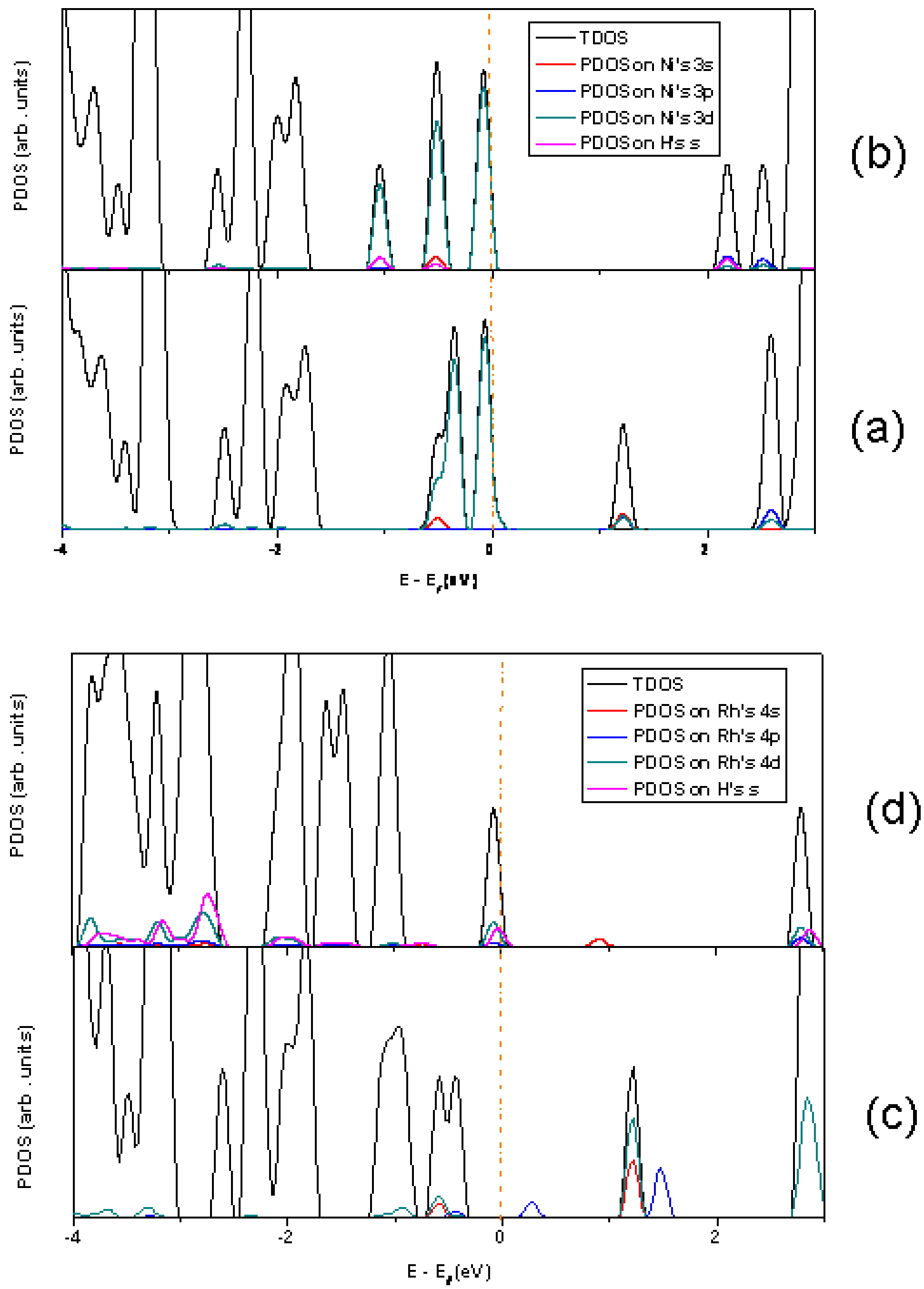
TDOS
PDOS on Ni's 3s
PDOS on Ni's 3p
PDOS on Ni's 3d
PDOS on H's s
(b)
(a)
PDOS (arb. units)
E - E$_F$ (eV)
TDOS
PDOS on Rh's 4s
PDOS on Rh's 4p
PDOS on Rh's 4d
PDOS on H's s
(d)
(c)
-4
-2
0
2

Figure 4

**Table: 1.** The absorption energy($E_{abs}$), the magnetic moment(μ), and the selected bond distances for the metal doped BN sheets with and without hydrogen adsorption.

| No. Hydrogen molecule | Ni | | | | | Rh | | | | |
|---|---|---|---|---|---|---|---|---|---|---|
| | $E_{abs}$ (eV) | μ (μB) | $d^{N-Ni}$ (Å) | $d_{avg}^{H-H}$ (Å) | $d_{avg}^{Ni-H}$ (Å) | $E_{abs}$ (eV) | μ (μB) | $d^{N-Rh}$ (Å) | $d_{avg}^{H-H}$ (Å) | $d_{avg}^{Rh-H}$ (Å) |
| 0 | – | 0.0 | 1.834 | – | – | – | 1.0 | 2.401 | – | – |
| 1 | 1.41 | 0.1 | 1.877 | 0.882 | 1.542 | 1.56 | 1.0 | 2.426 | 1.878 | 1.568 |
| 2 | 0.990 | 0.0 | 2.006 | 0.861 | 1.569 | 0.876 | 1.0 | 2.459 | 0.900 | 1.726 |
| 3 | 0.739 | 0.0 | 2.117 | 0.852 | 1.619 | 0.692 | 1.0 | 2.369 | 0.889 | 1.742 |